\documentclass{article}

\usepackage{PRIMEarxiv}
\usepackage{tabularx}
\usepackage[utf8]{inputenc} % allow utf-8 input
\usepackage[T1]{fontenc}    % use 8-bit T1 fonts
\usepackage{hyperref}       % hyperlinks
\usepackage{url}            % simple URL typesetting
\usepackage{booktabs}       % professional-quality tables
\usepackage{amsfonts}       % blackboard math symbols
\usepackage{nicefrac}       % compact symbols for 1/2, etc.
\usepackage{microtype}      % microtypography
\usepackage{lipsum}
\usepackage{fancyhdr}       % header
\usepackage{graphicx} 
\usepackage{amsmath} % graphics
\usepackage{float}
\usepackage{multirow}
\usepackage{graphicx}  % \resizebox
\usepackage{placeins} % \FloatBarrier
\graphicspath{{media/}}     % organize your images and other figures under media/ folder

\title{When to Trust the Map: Confidence-Aware LLM Routing for Automotive CVE-to-ATM Mapping\\
}

\author{

\begin{tabular}{c c c}
\begin{minipage}[t]{0.30\textwidth}\centering
\textbf{Heeyun Heo}\\
\normalfont School of Cybersecurity\\
Korea University\\
Seoul, Republic of Korea\\
\texttt{heeyun0724@korea.ac.kr}
\end{minipage}
& \hspace{2em} &
\begin{minipage}[t]{0.30\textwidth}\centering
\textbf{Sangmin Park}\\
\normalfont School of Cybersecurity\\
Korea University\\
Seoul, Republic of Korea\\
\texttt{psm9924@korea.ac.kr}
\end{minipage}
\\[2.5em]
\begin{minipage}[t]{0.30\textwidth}\centering
\textbf{Huy Kang Kim}\\
\normalfont School of Cybersecurity\\
Korea University\\
Seoul, Republic of Korea\\
\texttt{cenda@korea.ac.kr}
\end{minipage}
& \hspace{2em} &
\begin{minipage}[t]{0.30\textwidth}\centering
\textbf{Sanghoon Jeon}\\
\normalfont Department of Automobile and IT Convergence\\
Kookmin University\\
Seoul, Republic of Korea\\
\texttt{sh.jeon@kookmin.ac.kr}
\end{minipage}
\end{tabular}
}

\begin{document}
\maketitle

\begin{abstract}
Public CVE descriptions report the technical conditions and impact of vulnerabilities, whereas the Auto-ISAC Automotive Threat Matrix (ATM) expresses an adversary's tactics and techniques. Because the two representations are not directly aligned, incorrect automated mappings in safety-critical environments may distort threat interpretation and mitigation prioritization, motivating a confidence-aware approach that distinguishes auto-confirmable mappings from uncertain cases. This paper reformulates automotive CVE-to-ATM mapping as a selective automation problem. The proposed framework generates candidate mappings via hierarchical in-context learning, then fuses self-consistency and LLM-based evidence verification signals into a calibrated meta-model. The resulting confidence score $\hat{p}$ routes each candidate into AUTO, REVIEW, or HOLD. On the evaluation set, the proposed system substantially improved candidate-set precision at matched recall over a Flat zero-shot GPT-5.2 baseline. In the High-Confidence operating mode, the AUTO tier achieved a precision of $0.878$, more than double the candidate-set base rate, and the calibrated $\hat{p}$ achieved an AUROC of $0.868$ in distinguishing correct from incorrect candidates. These results show that the framework can support selective automation by isolating auto-confirmable mappings from those requiring analyst review.
\end{abstract}

% keywords can be removed

\section{Introduction}
\label{sec:intro}
As the automotive industry transitions toward software-defined vehicles (SDVs), vehicle functions are increasingly centralized in software, and external connectivity extends across cloud services, backend infrastructure, over-the-air (OTA) updates, and in-vehicle networks. This shift has substantially enlarged the attack surface~\cite{vicone2025}. 

Against this landscape, UN R155 requires manufacturers to establish a Cybersecurity Management System (CSMS) that monitors, detects, and responds to cyberattacks, cyber threats, and vulnerabilities, keeps risk assessments current, and mitigates relevant threats within a reasonable timeframe~\cite{unr155}. ISO/SAE 21434 complements this requirement through lifecycle-oriented cybersecurity engineering and TARA methods for road-vehicle E/E systems~\cite{iso21434}. However, CVEs from external threat intelligence mainly describe technical vulnerabilities, whereas automotive security practice frames threats as threat scenarios, attack paths, and vehicle-function impacts. Therefore, CVEs must be reinterpreted in terms of adversary objectives and attack behaviors before they can support practical security analysis. This reinterpretation can feed TARA updates, risk prioritization, and incident-response workflows, so incorrect mappings may affect downstream security decisions. Therefore, the mapping system should expose whether each tactic--technique pair is reliable enough for automatic adoption or requires analyst review.

Beyond confidence, such reinterpretation requires a domain-specific common threat language for automotive cybersecurity. General-purpose frameworks such as MITRE ATT\&CK provide useful threat knowledge bases, but their coverage of automotive-specific attack contexts such as CAN injection, OTA update compromise, and V2X spoofing remains limited~\cite{sommer2024combining}. To address this automotive coverage gap, Auto-ISAC released the Automotive Threat Matrix (ATM) in 2024, providing an automotive-domain tactic--technique taxonomy based on validated vehicle attack cases and security research. ATM was proposed as a common language for threat intelligence classification, risk assessment, incident response, and vulnerability management~\cite{autoIsacAtm2024}. In such frameworks, adversarial behavior is organized through tactics and techniques. A \textit{tactic} represents the adversary's high-level objective, or why an action is performed, whereas a \textit{technique} describes how that objective is achieved through a specific method.

However, manual analysis limits scalability and consistency. Mapping outcomes vary across analyst judgments, and interpreting each CVE against ATM definitions becomes an operational bottleneck for subsequent risk assessment. This burden grows with the increasing volume of automotive vulnerabilities. Consequently, a dependable automated mapping capability is required to translate CVEs into ATM tactic--technique pairs.

% Paragraphs 6 and 7 integrated: LLM potential and limitations

Large Language Models (LLMs) are natural candidates for such automation, as they can interpret unstructured CVE descriptions at the semantic level even with limited labeled data. In particular, they can leverage ATM definitions through in-context learning (ICL), which infers task patterns from prompt examples at inference time. However, automotive cybersecurity is a \emph{safety-critical} domain tied to vehicle functions, driver safety, supply-chain response, and regulatory reporting, where LLM errors may lead to incorrect threat prioritization or inappropriate TARA updates. Moreover, prior studies reported that LLMs suffer from hallucination, generation inconsistency, and limited self-verification \cite{ji2023survey}. Therefore, a single LLM output cannot meet the reliability level required for safety-critical automation.

To address this challenge, we formalize CVE-to-ATM mapping as a selective automation problem. The proposed framework integrates generation stability, evidence consistency, and confidence calibration to estimate the automation feasibility of each candidate. Based on this estimate, candidates are routed into three tiers: AUTO (auto-confirmation), REVIEW (analyst review), and HOLD. This routing establishes an explicit boundary between automated decisions and human oversight.

Based on this formulation, the main contributions of this paper are as follows:
\begin{itemize}

\item \textbf{Confidence-aware formulation of CVE-to-ATM mapping}: 
We formalize automotive CVE-to-ATM mapping as a selective automation problem in safety-critical environments.

\item \textbf{Multi-signal three-tier routing framework}: 
We propose a framework that generates candidate mappings through hierarchical in-context learning, verifies them with self-consistency and evidence-grounded judge signals, and combines these signals through a calibrated meta-model that drives the three-tier routing. Each routing decision carries its calibrated confidence and underlying signals, providing the decision traces required for CSMS audit and TARA update workflows.

\item \textbf{Automotive CVE-to-ATM benchmark dataset}: 
We construct a benchmark dataset of 269 automotive CVEs with 483 expert-annotated tactic--technique pairs.

\end{itemize}
The remainder of this paper is organized as follows. Section~\ref{sec:related} reviews related work, and Sections~\ref{sec:problem_dataset}--\ref{sec:deployment} present the proposed framework that formulates CVE-to-ATM mapping as a selective automation problem with three-tier human-in-the-loop routing. Section~\ref{sec:result} reports experimental results, and Section~\ref{sec:discussion} discusses practical implications for automotive cybersecurity.

\section{Related Work}
\label{sec:related}
\subsection{Automated CTI/CVE-to-TTP Mapping}

Two lines of prior work are adjacent to this study. First, ISO/SAE 21434's TARA process has motivated automation of internal threat modeling steps: AI has been introduced into STRIDE-based threat identification~\cite{ahmad2025ai}, and component-specific LLMs have been proposed for attack tree generation and risk assessment~\cite{yang2025defenseweaver}. These efforts focused on the internal stages of TARA and did not address how continuously reported CVE descriptions can be aligned with an automotive-specific threat taxonomy.

Second, prior efforts to extract Tactics, Techniques, and Procedures (TTPs) from unstructured cyber threat intelligence (CTI) evolved from ontology-based approaches~\cite{husari2017ttpdrill} to automated classification methods based on MITRE ATT\&CK~\cite{branescu2024cveTactics}. However, recent LLM benchmark studies report that despite their reasoning flexibility, LLMs remain unreliable for standalone operation due to hallucination and technical inaccuracies~\cite{alam2024ctibench,buchel2025sok}. This limitation is more critical in the automotive domain, where safety requirements exceed those of enterprise IT.

Recently, Scarano \textit{et al.} evaluated LLMs' automotive cybersecurity knowledge against the Auto-ISAC framework. They confirmed the potential of LLMs and called for a more fine-grained evaluation framework that separates model errors from cases requiring human expertise~\cite{scarano2025assessing}. This study addresses the gap between these two lines: it extends Scarano \textit{et al.}'s direction beyond knowledge assessment to CVE-to-ATM structuring, providing the missing external-vulnerability-to-taxonomy step that TARA-internal automation does not cover.

\subsection{Confidence Estimation and Deferral for LLM Outputs}
LLMs are increasingly used for security analysis tasks, but prior studies consistently reported that a single inference from one model cannot guarantee output reliability. Studies on LLM uncertainty estimation reported that LLMs are overconfident when verbalizing their confidence, and AUROC for distinguishing correct from incorrect answers remains limited even under prompting strategies or multi-response consistency calibration~\cite{xiong2024llmsuncertainty}. LLM-as-a-Judge, in which one LLM evaluates another's output, has also been reported as insufficient when relying on a single judgment~\cite{zheng2023judgingllmasajudgemtbenchchatbot}.

These limitations indicate the need for a multi-signal approach that independently collects and combines generation, verification, and calibration signals. General directions for managing model uncertainty, including learning-to-defer~\cite{madras2018predict} and conformal prediction~\cite{angelopoulos2022gentleintroductionconformalprediction}, have recently been extended to LLM-based systems, but existing studies have primarily addressed general-domain tasks such as question answering, classification, and summarization. Calibrated three-tier routing that integrates generation stability with evidence-grounded verification remains insufficiently explored in safety-critical domains such as automotive cybersecurity.

\section{Problem Formulation and Dataset}
\label{sec:problem_dataset}
\subsection{Problem Definition}
\label{sec:problem_def}
Given a single CVE description, the system outputs a set of ATM tactic--technique pairs and a routing decision for each pair. CVE-to-ATM mapping requires semantic reasoning to interpret adversarial behavior in a vehicle environment from a limited vulnerability description. The proposed framework therefore estimates a calibrated confidence for each candidate and routes it accordingly.

\paragraph{Input.}
The input is a single CVE description $D \in \mathcal{T}$, where $\mathcal{T}$ is the space of CVE texts. At inference time, the system does not use external threat intelligence, vendor advisories, proof-of-concept reports, or web search results.

\paragraph{Output.}
For each CVE, the system produces a set of $K(D)$ output triples:
\begin{equation}
\mathcal{O}(D)=\left\{\left((t_a^{(i)}, t_e^{(i)}),\; \hat{p}^{(i)},\; 
\mathrm{tier}^{(i)}\right)\right\}_{i=1}^{K(D)}
\label{eq:output}
\end{equation}
Here, $t_a^{(i)} \in \mathcal{A}$ is an ATM tactic from the set of all tactics $\mathcal{A}$, $t_e^{(i)} \in \mathcal{E}(t_a^{(i)})$ is a technique drawn from $\mathcal{E}(t_a^{(i)})$ — the set of techniques associated with tactic $t_a^{(i)}$, $\hat{p}^{(i)} \in [0,1]$ is the calibrated confidence, and $\mathrm{tier}^{(i)} \in \{\text{AUTO}, \text{REVIEW}, \text{HOLD}\}$ is the human-in-the-loop routing decision. Pairs from the same CVE may be assigned to different tiers according to their respective $\hat{p}$ values. 

\subsection{Dataset Construction}
\label{sec:dataset}
Because no public dataset exists for mapping CVE descriptions to ATM tactic--technique pairs, we construct an expert-annotated benchmark from publicly available automotive security resources.

\paragraph{Data sources.}
We collected automotive vulnerability data from two primary public sources, supplemented by additional automotive-related CVEs gathered from public vulnerability databases.
VicOne Automotive Zero-day Advisories~\cite{vicone2026zeroday} provide recent advisories and zero-day information. The Automotive Threat Database (AutomotiveTD)~\cite{jayaratne2023automotivetd} covers threat cases across diverse vehicle systems such as ECUs, telematics, and V2X. We deduplicated the collected data by CVE ID and integrated them after normalizing key fields such as the CVE ID, description, and publication date. The final dataset comprises 269 CVEs and 483 tactic--technique pairs, and 51\% of the CVEs carry multiple labels.

\paragraph{Annotation protocol.}
Four experts in automotive cybersecurity performed independent per-pair annotation after reviewing the official ATM definitions, consulting public external evidence (vendor advisories, proof-of-concept reports, security research blogs) when available and otherwise relying on the CVE description alone. Labels were restricted to directly verifiable attack stages, excluding indirect secondary effects and speculative extensions. Disagreements were resolved through consensus by matching evidence phrases against the official ATM definitions. CVEs without consensus or outside the ATM framework were excluded as unmappable. Final labels correspond to pairs $(t_a, t_e)$ from the 14 tactics and 75 techniques of the ATM schema.

\paragraph{Split design.}
To improve evaluation reliability on this small dataset, we applied two split-design principles. First, to prevent data contamination from near-duplicate descriptions of the same underlying vulnerability, we vectorized all CVE descriptions using TF-IDF, grouped pairs with similarity above 0.85, and ensured that no group spanned multiple splits. Second, while preserving group integrity, we partitioned the groups by CVE publication year (2010--2025). Earlier groups formed the Exemplar Pool (149 CVEs), from which a subset of examples is dynamically retrieved per query during ICL. More recent groups were assigned to the Development Set (62 CVEs) for system optimization and the Evaluation Set (58 CVEs) for final performance measurement. As a result, 100\% of the Evaluation Set consists of CVEs published in 2024--2025, and no 2025 CVE appears in the Exemplar Pool. Group integrity occasionally forced adjacent-year overlaps across splits, but the split still approximates temporal generalization.

\section{Methodology}
\label{sec:methodology}
\subsection{System Overview}
\label{sec:overview}
\begin{figure}[H]
\centering
\includegraphics[width=\linewidth]{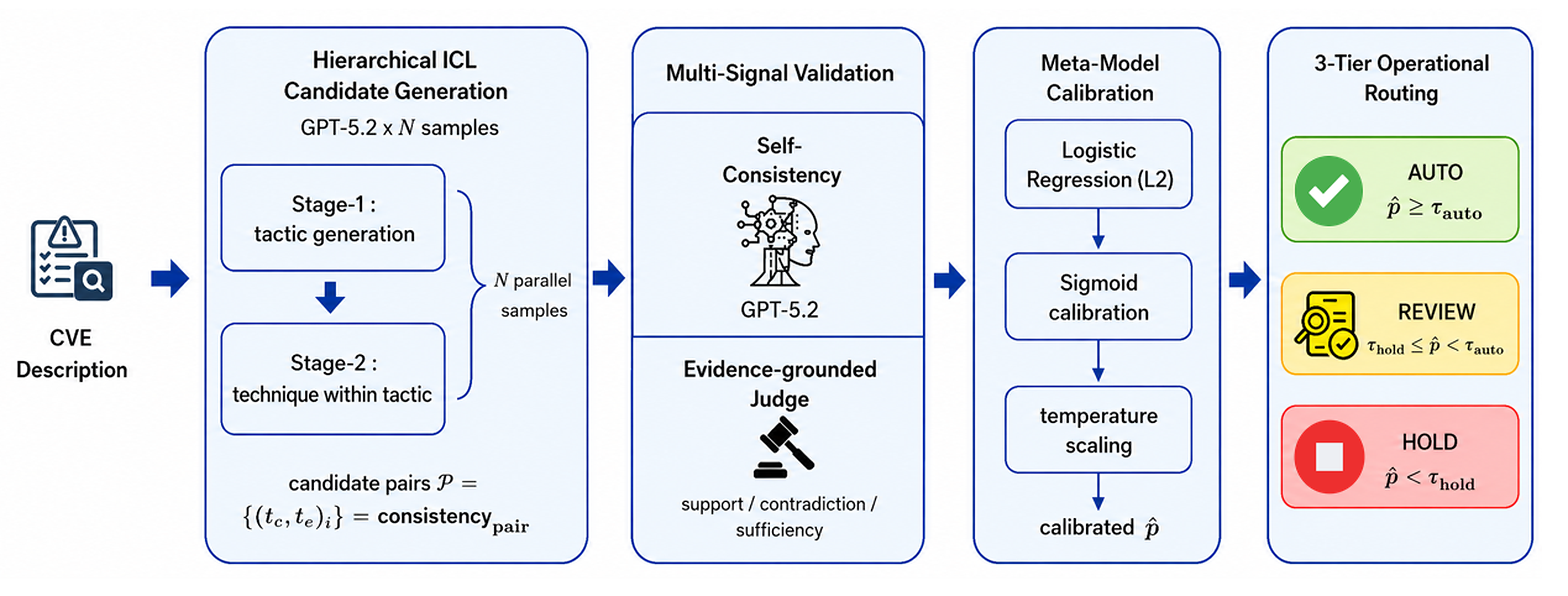}
\caption{Overview of the proposed confidence-aware CVE-to-ATM mapping pipeline.}
\label{fig:pipeline}
\end{figure}

The proposed system consists of four stages (Fig.~\ref{fig:pipeline}): candidate generation, multi-signal verification, confidence calibration, and operational routing.

\subsection{Two-stage Candidate Generation}
\label{sec:candidate}
The proposed system leverages the hierarchical ATM structure by decomposing the mapping task into two stages: (1) identifying relevant tactic candidates and (2) predicting techniques conditioned on each identified tactic. Both stages use ICL with examples drawn from the Exemplar Pool, without additional model training. We use OpenAI GPT-5.2 (\texttt{reasoning effort = high}) as the backbone model. A fixed system prompt frames the model as an expert in automotive cybersecurity and the Auto-ISAC ATM, aligning its outputs with the target ontology.

\paragraph{Stage 1: Tactic prediction.}
The first stage uses a prompt consisting of (i) system instructions summarizing the ATM tactic definitions, (ii) CVE--tactic examples drawn from the Exemplar Pool, and (iii) the description of the target CVE. A function-calling schema constrains the model to select from the 14 valid tactic IDs, preventing ontology-violating hallucinations by construction.

\paragraph{Stage 2: Tactic-conditioned technique prediction.}
For each tactic candidate from Stage 1, the second stage dynamically constructs a schema that includes only the techniques associated with that tactic, blocking pairs that violate the tactic--technique hierarchy by construction. ICL examples are drawn only from Exemplar Pool entries labeled with the target tactic, from which the top $k{=}8$ cases most semantically similar to the input CVE description are retrieved via $k$-nearest-neighbor ICL (kNN-ICL). This retrieval focuses the model on tactic-specific examples that are semantically close to the input CVE, while keeping prompt length bounded.

\paragraph{Repeated inference and pair-level consistency.}
To mitigate response inconsistency across repeated LLM calls, Stage 1 is invoked $N=5$ times per CVE, and Stage 2 is invoked $N=5$ times per tactic candidate from Stage 1. The consistency $c(\cdot)$ of a tactic $t_a$ or a conditional technique $t_e \mid t_a$ is defined as the proportion of generations in which it appears, and pair-level consistency is their product:
\begin{equation}
\mathrm{consistency}_{\mathrm{pair}}(t_a,t_e)=c(t_a)\cdot c(t_e \mid t_a).
\end{equation}
The candidate set $\mathcal{P}(D)$ retains all valid pairs observed at least once, since a majority-voting filter could discard low-frequency but potentially valid threat mappings. Incorrect mappings are subsequently pruned by the multi-signal verification stage (\S\ref{sec:validation}), regardless of their generation frequency.

\subsection{Multi-Signal Verification}
\label{sec:validation}
Because the candidate set $\mathcal{P}(D)$ relies solely on generative output, generation frequency alone cannot distinguish a repeatedly generated incorrect pair from a correct one. To address this, we combine two complementary signals: \textbf{pair-level consistency} (\S\ref{sec:candidate}), which captures generation stability, and \textbf{evidence-grounded verification}, which checks factual alignment between CVE-extracted evidence and candidate technique definitions through a separate verification call. The verification stage adapts the LLM-as-a-Judge paradigm by grounding judgments in evidence explicitly extracted from the CVE rather than the verifier's prior knowledge. We further anonymize candidates before judging and decompose the verification signal into three distinct axes. Disagreement between the two signals exposes pairs that are frequently generated yet unsupported by CVE evidence.

\paragraph{Evidence-grounded LLM verification.}
A separate verification call, acting as an evidence-grounded judge, reassesses each candidate pair against the facts in the CVE text. The system first extracts six structured elements informed by ISO/SAE 21434 TARA—covering the affected component, attack interface, security consequence, attack precondition, protocol or bus, and vehicle impact—along with supporting evidence spans. Any element not explicitly stated is marked as unknown, which constrains the verifier to evidence-only reasoning. After anonymizing the candidates, the verifier outputs three scores in $[0,1]$ for each pair: \textit{support}, \textit{contradiction}, and \textit{sufficiency}. This separation prevents low sufficiency and high contradiction from yielding identical low values, which improves the interpretability of the verification signal.

\subsection{Confidence Integration and Calibration}
\label{sec:meta}
We train a \emph{meta-model} on the Development Set that integrates the signals from \S\ref{sec:validation} into a single calibrated confidence $\hat{p}$ for each candidate pair.

\paragraph{Feature vector.}
Each candidate pair $(t_a, t_e) \in \mathcal{P}(D)$ is encoded as a six-dimensional feature vector $\mathbf{x} = [x_1, \ldots, x_6]^\top$. Here, $x_1 =\mathrm{consistency}_{\mathrm{pair}}$ 
captures generation stability, and $x_2, x_3, x_4$ correspond to the verifier's three scores: \textit{support}, \textit{contradiction}, and \textit{sufficiency}. To capture conjunctive evidence, we add two interaction terms:
\begin{equation}
x_5 = x_1 \cdot x_2, \quad x_6 = x_1 \cdot x_4.
\end{equation}
These terms enable the meta-model to increase confidence only when generation stability ($x_1$) and verification-based factual support ($x_2, x_4$) are jointly high.

\paragraph{Meta-model and calibration.}
We adopt L2-regularized logistic regression as the meta-model for its interpretability and stability on small-scale data. The model outputs $\hat{p}$, the estimated probability that a candidate mapping is correct. To support threshold-based routing, we apply \textit{post hoc} calibration on the Development Set in two stages. Sigmoid calibration first corrects the overall scale mismatch. Temperature scaling then addresses residual overconfidence or underconfidence by selecting $T$ that minimizes validation negative log-likelihood (NLL) via grid search. The temperature scaling step intervenes only when needed, avoiding unconditional distortion of 
the score distribution. The resulting $\hat{p} \in [0,1]$ drives the AUTO, REVIEW, and HOLD routing in \S\ref{sec:routing}. 

\section{Operational Routing and Deployment}
\label{sec:deployment}
\subsection{Three-Tier Human-in-the-Loop Routing}
\label{sec:routing}
Building on the calibrated $\hat{p}$ produced by the meta-model in \S\ref{sec:meta}, we apply a three-tier human-in-the-loop (HITL) routing policy. The policy adjusts the balance between automation and analyst review by changing thresholds alone, without retraining the meta-model.

\paragraph{Operational modes.}
AUTO candidates have sufficient calibrated confidence for automatic confirmation. REVIEW candidates require analyst inspection with supporting confidence and evidence. HOLD candidates are deferred until additional evidence or reanalysis is available. Routing operates at the pair level, so candidates from the same CVE may receive different tiers. 

\paragraph{Mode selection.}
To map the precision--coverage trade-off available under the calibrated $\hat{p}$, we swept $\tau_{\text{auto}}$ across the Development-Set $\hat{p}$ distribution. The three operational modes share the same $\tau_{\text{hold}}$ and differ only in $\tau_{\text{auto}}$. From this sweep, we select three operating points that span distinct regions of the trade-off curve and report them as named deployment modes. The corresponding thresholds and calibration parameters are then applied to the Evaluation Set without re-selection.

\begin{itemize}
  \item \textbf{High-Confidence mode}: a conservative operating point selected from the high-precision region of the Development sweep. This mode targets autonomous CSMS workflows in which AUTO confirmations enter the vulnerability registry without further analyst review, where false positives carry a high operational cost.
  \item \textbf{Balanced mode}: an operating point from the mid-region of the sweep, retaining substantial recall while keeping AUTO precision well above the random baseline.
  \item \textbf{High-Coverage mode:} an exploratory operating point from the high-recall region of the sweep, used to assess the volume of correct mappings the system can route at low selectivity.
\end{itemize}

\paragraph{Routing policy.}
The tier assignment of each candidate pair is given by:
\begin{equation}
(t_a, t_e) \longrightarrow 
\begin{cases}
\text{AUTO}   & \text{if } \hat{p} \geq \tau_{\text{auto}}, \\
\text{REVIEW} & \text{if } \tau_{\text{hold}} \leq \hat{p} < \tau_{\text{auto}}, \\
\text{HOLD}   & \text{if } \hat{p} < \tau_{\text{hold}}.
\end{cases}
\end{equation}

\section{Results}
\label{sec:result}
\subsection{Main Results}
\label{sec:main_results}

\paragraph{Experimental setup.}
We conducted the evaluation on the Development Set (62 CVEs, 102 gold tactic--technique pairs) and the Evaluation Set (58 CVEs, 98 gold pairs). The data split, meta-model training, and threshold selection follow the design described in \S\ref{sec:dataset} through \S\ref{sec:routing}. For the comparative evaluation in Table~\ref{tab:main_end2end}, all systems share the same 98-gold-pair denominator. We compare against three methodologically representative baselines:

\begin{itemize}
    \item \textbf{(A) TF-IDF + Linear Classifier:}
    A classical supervised baseline using TF-IDF features and a linear classifier.
    \item \textbf{(B) Flat Zero-shot GPT-5.2:}
    A zero-shot LLM baseline that directly predicts ATM pairs using a single prompt, without structured decomposition or exemplar retrieval.
    \item \textbf{(C) Frozen SecureBERT Retrieval:}
    A retrieval baseline that, for each Evaluation CVE, retrieves the most semantically similar CVEs from the Exemplar Pool using frozen SecureBERT and uses their gold tactic--technique labels as candidate mappings.
\end{itemize}

\paragraph{Comparative Performance.}
As shown in Table~\ref{tab:main_end2end}, \textit{Ours -- all candidates} outperformed all three baselines on every pair-level metric. Against the Flat Zero-shot GPT-5.2 baseline in particular, it matched recall while raising precision by $+0.118$ and reducing average predictions from $4.55$ to $3.19$. This indicates that hierarchical decomposition and semantic exemplar retrieval improve candidate-set quality rather than merely increasing candidate count.

\begin{table}[!htbp]
\centering
\footnotesize
\caption{End-to-end comparison on the Evaluation Set. All systems report raw candidate predictions before routing is applied. P and R are micro-averaged over all predicted pairs against 98 gold pairs. MRR and R@5 are computed per CVE: pair-level over $(tactic, technique)$ rankings, tactic-level over the same rankings deduplicated per tactic. Tactic-level R@5 is upper-bounded by $|\mathrm{tactics}|{=}14$. Best in bold.}
\label{tab:main_end2end}
\resizebox{\linewidth}{!}{%
\begin{tabular}{lcccccc}
\toprule
Configuration & P & R & F1 & MRR & R@5 & AvgPred \\
\midrule
\multicolumn{7}{l}{\textit{Pair-level evaluation: fine-grained $(tactic, technique)$ pairs}} \\
\midrule
(A) TF-IDF + Linear              & 0.302 & 0.459 & 0.364 & 0.477 
& 0.592 & 2.57 \\
(B) Flat Zero-shot GPT-5.2       & 0.277 & 0.745 & 0.403 & 0.601 
& 0.727 & 4.55 \\
(C) Frozen SecureBERT Retrieval  & 0.246 & 0.306 & 0.273 & 0.397 
& 0.463 & 2.10 \\
\midrule
Ours -- all candidates           & \textbf{0.395} & \textbf{0.745} 
& \textbf{0.516} & \textbf{0.716} & \textbf{0.783} & 3.19 \\
\midrule
\multicolumn{7}{l}{\textit{Tactic-level evaluation: coarse-grained labels for comparability with prior CVE-to-TTP studies}} \\
\midrule
(A) TF-IDF + Linear              & 0.425 & 0.543 & 0.477 & 0.588 
& 0.813 & 2.07 \\
(B) Flat Zero-shot GPT-5.2       & 0.434 & 0.872 & 0.580 & 0.721 
& 0.897 & 3.26 \\
(C) Frozen SecureBERT Retrieval  & 0.333 & 0.415 & 0.370 & 0.513 
& 0.675 & 2.02 \\
\midrule
Ours -- all candidates           & \textbf{0.529} & \textbf{0.883} 
& \textbf{0.661} & \textbf{0.805} & \textbf{0.917} & 2.71 \\
\bottomrule
\end{tabular}%
}
\end{table}

\FloatBarrier

\paragraph{Routing results.}
The proposed system generated 185 pair-level candidates covering 73 of 98 gold pairs; the remaining 25 lie outside the candidate set. Table~\ref{tab:routing} reports the routing outcome. The three modes share the same calibrated $\hat{p}$ and differ only in the AUTO threshold $\tau_{\text{auto}}$. Within the candidate set, the base rate of correct pairs is approximately $0.395$, which sets the reference point against which each $\tau_{\text{auto}}$ is interpreted. The three operating points correspond to $\tau_{\text{auto}} = 0.21$ for High-Coverage, $0.38$ for Balanced, and $0.7431$ for High-Confidence, yielding AUTO precisions above, near twice, and more than twice this base rate. Each threshold is selected from a distinct region of the Development Set $\hat{p}$ sweep described in \S\ref{sec:routing}.

\begin{table}[!htbp]
\centering
\footnotesize
\caption{Pair-level routing on the Evaluation Set. For each tier, 
$N$ is the number of candidates assigned to it, $\text{Share}=N/185$, and Correct is the number of correct candidates in the tier. Tier P denotes within-tier precision $\text{Correct}/N$. R, F1, and AvgPred summarize the AUTO row of each mode, where R and F1 use 98 gold pairs as denominator (directly comparable to Table~\ref{tab:main_end2end}) and AvgPred reports the average number of AUTO predictions per CVE. Best in bold; AvgPred is descriptive only.}
\label{tab:routing}
\resizebox{\linewidth}{!}{%
\begin{tabular}{llcccc|ccc}
\toprule
Mode & Tier & $N$ & Share & Correct & Tier P & R & F1 & AvgPred \\
\midrule
\multicolumn{9}{l}{\textit{Total 185 candidates, reachable gold = 
73, total gold = 98}} \\
\midrule
\multirow{3}{*}{High-Coverage ($\tau{=}0.21$)}
  & AUTO   & 125 & 0.676 & 67 & 0.536 & \textbf{0.684} & 0.601 & 
2.16 \\
  & REVIEW &  53 & 0.286 &  6 & 0.113 & --- & --- & --- \\
  & HOLD   &   7 & 0.038 &  0 & 0.000 & --- & --- & --- \\
\midrule
\multirow{3}{*}{Balanced ($\tau{=}0.38$)}
  & AUTO   &  83 & 0.449 & 60 & 0.723 & 0.612 & \textbf{0.663} & 
1.43 \\
  & REVIEW &  95 & 0.514 & 13 & 0.137 & --- & --- & --- \\
  & HOLD   &   7 & 0.038 &  0 & 0.000 & --- & --- & --- \\
\midrule
\multirow{3}{*}{High-Confidence ($\tau{=}0.7431$)}
  & AUTO   &  49 & 0.265 & 43 & \textbf{0.878} & 0.439 & 0.585 & 
0.84 \\
  & REVIEW & 129 & 0.697 & 30 & 0.233 & --- & --- & --- \\
  & HOLD   &   7 & 0.038 &  0 & 0.000 & --- & --- & --- \\
\bottomrule
\end{tabular}%
}
\end{table}

The Balanced threshold $\tau_{\text{auto}} = 0.38$ keeps recall substantial at the cost of per-pair precision, making it appropriate for analyst-audited workflows. For workflows where AUTO confirmations are committed without further review, High-Confidence is the intended operating point.

All seven HOLD candidates with $\hat{p} < \tau_{\text{hold}}=0.10$ 
were incorrect, which confirms that HOLD effectively separates low-confidence invalid candidates. As $\tau_{\text{auto}}$ increases, the AUTO set shrinks while within-tier precision rises. High-Coverage supports broad initial screening at low selectivity. High-Confidence achieves an AUTO precision of $0.878$, over $2\times$ the base rate, providing the quantitative basis for its use in autonomous CSMS deployment. The REVIEW set correspondingly grows from $53$ to $129$ candidates as uncertain candidates are deferred to analyst review. Within REVIEW, candidates remain ranked by $\hat{p}$, with within-tier AUROC of $0.74$ for High-Confidence, allowing analysts to prioritize top-ranked items within the tier.

\paragraph{Analysis of the 25 missed gold pairs.}
The 25 missed gold pairs are dominated by two recurring patterns that together account for 22 of 25 misses (88\%). First, 14 misses (56\%) are Stage 2 failures, with 10 in \textit{Execution} (6/6) and \textit{Initial Access} (4/4), in which the correct tactic was identified but semantically adjacent techniques were not distinguishable from the CVE description alone. Second, 8 misses (32\%) are Stage 1 failures, in which auxiliary tactics whose strategic role is implicit in CVE text (\textit{Defense Evasion}, \textit{Command and Control}) were under-predicted. Both patterns reflect a structural gap: CVE descriptions report technical defects, while ATM annotations encode adversary intent.

\subsection{Ablation Study}
\label{sec:ablation}
To assess how self-consistency and evidence-grounded judge signals contribute to the meta-model, we retrained it after removing each signal group and its interaction terms (Table~\ref{tab:ablation}). 
Removing the judge signals reduced the Area Under the ROC Curve (AUROC) from $0.868$ to $0.705$, identifying the evidence-grounded judge as the primary driver of discriminative ranking. Removing the consistency signal lowered AUROC to $0.851$, while the Expected Calibration Error (ECE) improved from $0.057$ to $0.054$.

The feature coefficients confirm this picture. The highest-weighted features are \textit{judge\_support} ($+0.485$), \textit{judge\_sufficiency} ($+0.391$), and the two interaction terms \textit{consistency$\times$support} ($+0.335$) and \textit{consistency$\times$sufficiency} ($+0.243$). Positive interaction coefficients show that judge signals contribute more when consistency is high. The direct coefficient of \textit{consistency\_pair} is small ($+0.082$), but consistency also enters the model through both interaction terms. Therefore, its total contribution is distributed across direct and interaction pathways rather than captured by the direct coefficient alone.

\begin{table}[H]
\centering
\footnotesize
\setlength{\abovecaptionskip}{4pt}
\setlength{\belowcaptionskip}{4pt}
\caption{Component ablation on the Evaluation Set. Full system results in the first row; bold marks the best value in each column. Lower ECE indicates better calibration; higher AUROC indicates better ranking discrimination.}
\label{tab:ablation}
\begin{tabular}{lcc}
\toprule
Configuration & ECE & AUROC \\
\midrule
Full (6 features)                   & 0.057          & \textbf{0.868} \\
\quad $-$ consistency (judge only)  & \textbf{0.054} & 0.851 \\
\quad $-$ judge (consistency only)  & 0.059          & 0.705 \\
\quad $-$ calibration               & 0.071          & \textbf{0.868} \\
\bottomrule
\end{tabular}
\end{table}
\vspace{-0.8em}

\subsection{Stability and Sensitivity Analysis}
\paragraph{Bootstrap stability.}
We performed $B{=}1000$ CVE-level bootstrap resamples on the Evaluation split ($N{=}58$). The calibrated $\hat{p}$ achieved AUROC $0.868$ (95\% CI $[0.815, 0.916]$), exceeding the consistency-only baseline. A paired bootstrap confirmed $\Delta\text{AUROC}=+0.163$ (CI $[+0.104, +0.229]$) over the consistency-only baseline (Table~\ref{tab:ablation}) and $\Delta F_1=+0.113$ (CI $[+0.056, +0.168]$) over the Flat Zero-shot GPT-5.2 baseline. Both intervals exclude zero. In the High-Confidence mode, AUTO precision reached $0.878$ (95\% CI $[0.778, 0.959]$).

\begin{figure}[!t]
\centering
\includegraphics[width=0.65\linewidth]
{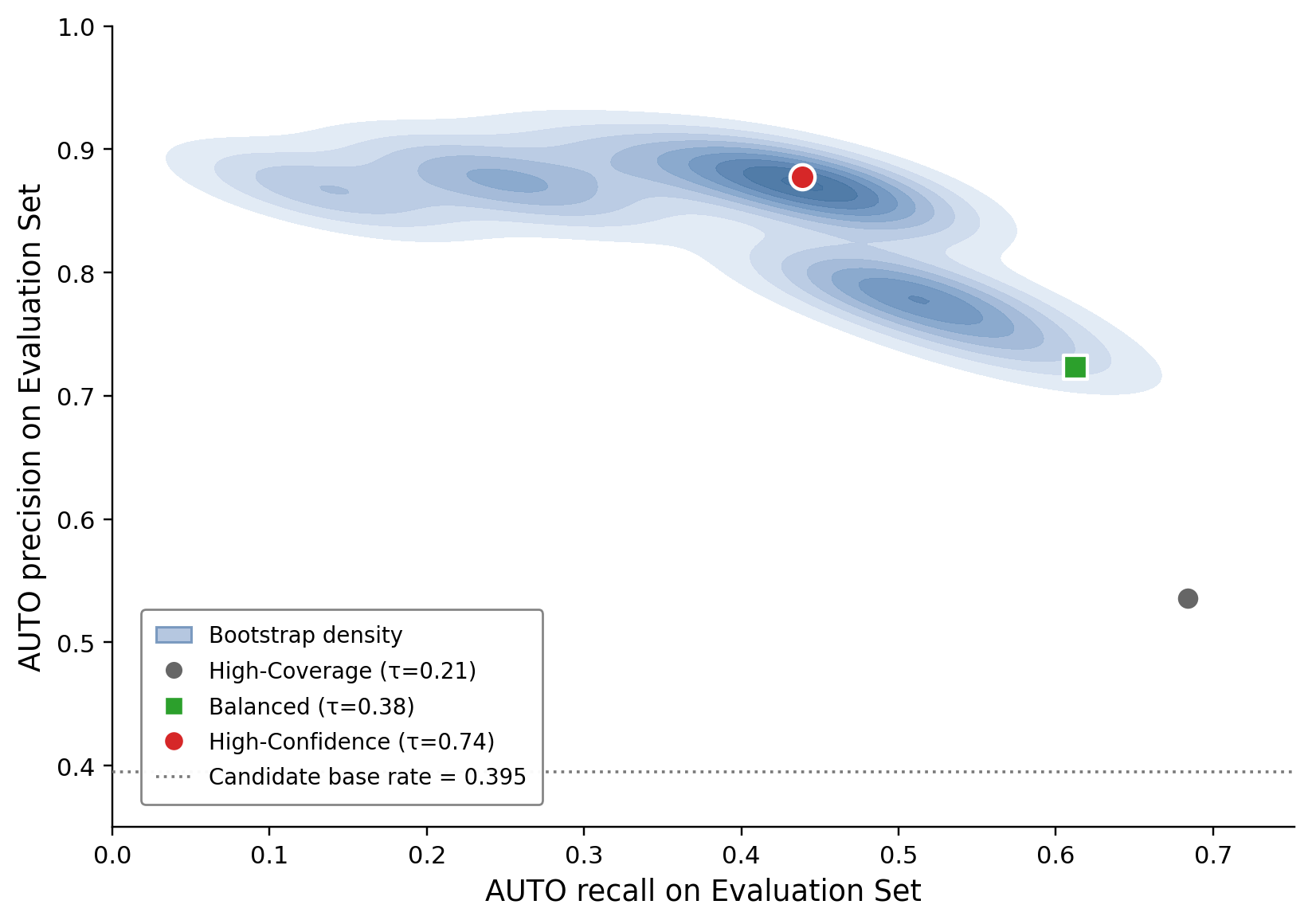}
\caption{Bootstrap precision--recall of the AUTO tier as
$\tau_{\text{auto}}$ varies across its 95\% CI $[0.428, 0.942]$
($B{=}1000$). AUTO precision stays within $0.78$--$0.90$, with the High-Confidence mode ($\tau{=}0.74$) inside the density and the High-Coverage mode ($\tau{=}0.21$) serving as a coverage anchor outside the CI. All three modes exceed the candidate base rate ($0.395$, dotted line).}
\label{fig:threshold_sensitivity}
\end{figure}

\paragraph{Threshold sensitivity.}
A bootstrap analysis on the Development set ($N{=}62$) showed that $\tau_{\text{auto}}$ in the High-Confidence mode had a 95\% CI of $[0.428, 0.942]$. To examine whether this wide CI affects deployment stability, we re-evaluated the AUTO tier on the Evaluation Set across $\tau_{\text{auto}}$ values resampled from this CI (Fig.~\ref{fig:threshold_sensitivity}).
The resulting AUTO precision remained within $0.78$--$0.90$ throughout the CI, and the High-Confidence operating point ($\tau{=}0.74$) fell within the high-density region of the bootstrap distribution. The wide $\tau_{\text{auto}}$ CI therefore reflects the precision--coverage trade-off available to the operator rather than instability of the routing decision.

\FloatBarrier
\section{Discussion}
\label{sec:discussion}
The central implication of this study is that automotive CVE-to-ATM mapping can be reformulated as a selective automation problem in which auto-confirmation and analyst review are separated according to confidence. On the Evaluation Set, the calibrated $\hat{p}$ achieved an AUROC of $0.868$ versus $0.705$ for the consistency-only configuration (Table~\ref{tab:ablation}), indicating that self-consistency alone is insufficient and that evidence-grounded judge signals are necessary for reliable confidence estimation. In the High-Confidence mode, the AUTO tier reached a precision of $0.878$, more than $2\times$ the candidate base rate of $0.395$, which indicates that the calibrated $\hat{p}$ separates auto-confirmable mappings from the broader candidate pool rather than merely reflecting an arbitrary cut-off. 

These properties have direct operational implications. The framework auto-confirms only sufficiently confident tactic--technique pairs and routes uncertain cases to analysts with calibrated $\hat{p}$ and evidence-grounded signals, preserving analyst attention for ambiguous cases while maintaining traceable decision evidence for cybersecurity work products required by ISO/SAE 21434 Clause~15~\cite{iso21434}. When the ATM ontology is updated~\cite{autoIsacAtm2024}, maintenance can be localized to the exemplar pool, routing thresholds, and lightweight meta-model.

This operational interpretation should be understood within the following scope. First, the evaluation was conducted on the current ATM ontology (14 tactics, 75 techniques) and 269 expert-annotated CVEs, so ontology revisions or OEM-specific taxonomies may require updating the ICL exemplars and retraining the meta-model. Second, both the candidate generation stage and the evidence-grounded judge use the same GPT-5.2 backbone. This same-backbone design may raise concerns about self-confirmation, but the framework reduces direct confirmation bias through evidence grounding, candidate anonymization, and decomposed judgment. The verifier is grounded in a CVE-only evidence packet, where unstated fields are explicitly marked as unknown, and it receives anonymized candidates without generation order or consistency scores. The verifier also outputs support, contradiction, and sufficiency separately. This separation distinguishes weak evidence from explicit conflict. The ablation in \S\ref{sec:ablation} supports this design choice: removing the judge reduced AUROC from $0.868$ to $0.705$. Nevertheless, cross-family validation remains necessary to test whether these safeguards generalize beyond a single backbone. Third, because the meta-model integrates judge signals as continuous features, some boundary cases with weak evidence support may still receive high confidence; a minimum-support condition or rule-based guardrail could mitigate this. Fourth, the effectiveness of analyst review in the REVIEW tier requires field study with security analysts.

\section{Conclusion}
\label{sec:conclusion}
This paper proposes a confidence-aware framework that formulates automotive CVE-to-ATM mapping as a three-tier routing problem over calibrated confidence. Each tactic--technique pair is assigned to the AUTO, REVIEW, or HOLD tier based on its $\hat{p}$. On the Evaluation Set, the calibrated $\hat{p}$ achieved an AUROC of $0.868$, exceeding the consistency-only baseline by $+0.163$, and the AUTO tier in the High-Confidence mode reached a precision of $0.878$, more than $2\times$ the candidate base rate. To the best of our knowledge, this is an early attempt to apply calibration-aware routing to the Auto-ISAC Automotive Threat Matrix.

By exposing the confidence underlying each mapping, the framework provides an explicit boundary between automated decisions and analyst review that automotive CSMS workflows can build on. Larger datasets and analyst field study in the REVIEW tier remain future work.

%Bibliography
\bibliographystyle{unsrt}  
\bibliography{references}

\end{document}